\documentclass[floatfix, reprint, amsmath,amssymb,aps, prx,unsorted, nobibnotes]{revtex4-2}
\usepackage{amsmath,amssymb,amsfonts}
\usepackage{amsthm}
\usepackage{physics}
\usepackage{graphicx}
\usepackage{dcolumn}
\usepackage{bm}
\usepackage{lipsum}
\usepackage{xcolor}
\usepackage{booktabs, makecell}
\usepackage{subcaption}

\begin{document}

\title{Quantum reservoir computing induced by controllable damping}
\author{Emanuele Ricci}
\author{Francesco Monzani}
\author{Luca Nigro}
\author{Enrico Prati}
\thanks{e-mail: enrico.prati@unimi.it}
\affiliation{Dipartimento di Fisica "Aldo Pontremoli", Università degli Studi di Milano, Via Celoria 16, 20133, Milan, Italy}

\begin{abstract}
Quantum reservoir computing has emerged as a promising machine learning paradigm for processing temporal data on near-term quantum devices, as it allows for exploiting the large computational capacity of the qubits without suffering from typical issues that occur when training a variational quantum circuit. In particular, quantum gate-based echo state networks have proven effective for learning when the evolution of the reservoir circuit is non-unital. Nonetheless, a method for ensuring a tunable and stable non-unital evolution of the circuit was still lacking. We propose an algorithm for inducing damping by applying a controlled rotation to each qubit in the reservoir. It enables tunable, circuit-level amplitude amplification of the zero state, maintaining the system away from the maximally mixed state and preventing information loss caused by repeated mid-circuit measurements. The algorithm is inherently stable over time as it can, in principle, process arbitrarily long input sequences, well beyond the coherence time of individual qubits, by inducing an arbitrary damping on each qubit. Moreover, we show that quantum correlations between qubits provide an improvement in terms of memory retention, underscoring the potential utility of employing a quantum system as a computational reservoir.  We demonstrate, through typical benchmarks for reservoir computing, that such an algorithm enables robust and scalable quantum random computing on fault-tolerant quantum hardware.
\end{abstract}

\maketitle

\section{Introduction}
Motivated by promising insights for quantum algorithms for machine learning, the challenge of implementing them on currently available quantum hardware
 -- currently limited by the inherent noise of noisy intermediate-scale quantum (NISQ) devices -- is gaining increasing attention~\cite {PRXQuantum.1.020101, rev.q.alg,prati2017quantum}. In particular, non-variational algorithms, such as quantum kernel methods \cite{schuld2021machine}, quantum principal component analysis \cite{lloyd2014quantum}, and extreme learning machine \cite{innocenti2023potential}, which do not require training of internal parameters of the quantum circuit, emerge as successful techniques since they inherently avoid the typical training problems \cite{barren,larocca2025barren}. Among them, quantum reservoir computing stands out as particularly effective in handling sequential data processing for time series analysis, forecasting of chaotic dynamics, anomaly detection, and classification tasks \cite{qrc_opp}. We present an algorithm for gate-based quantum reservoir computing and discuss the essential features for its effective implementation. Next, we propose an explicit circuital embodiment. 
Reservoir computing is a well-established supervised machine learning algorithm that employs a fixed neural network to process sequential data, originally stemming from the echo state network~\cite{jaeger2001short} and the liquid state machine~\cite{maass.nat, maass} models. It has emerged as a paramount example of unconventional computing deployment \cite{rc.naka, monz_uni}, with diverse physical implementations~\cite{tanaka2019recent}. Among them, many quantum systems -- associated with their exponentially large computational capacity --
has been harnessed for reservoir computing~\cite{grollier, senanian2024microwave, franceschetto, ghosh2019quantum, ghosh2021realising, gotting2023exploring, ph1, ph2, neutral.atoms, boso1, nokkala2021gaussian, Spagnolo2022, pfeffer, magri, dudas2023quantum, yama}. Focusing on superconducting qubits embodiments of quantum reservoir computing, besides early implementations~\cite{chen2, molteni, suzuki2022natural}, they have recently been empowered by the possibility of mid-circuit measurements~\cite{kubota2023temporal, hu2024overcoming, xiong2025fundamental}. Nonetheless, the repeated measurements of the qubit register progressively steer the quantum system toward the maximally mixed state, thereby reducing the expressivity of the network and hindering its ability to distinguish different input values \cite{qrc}. Thus, the loss of information should be mitigated by inducing a partial purification of the state of the quantum network. In previous works \cite{noiyqrcmonz, Domingo2023, zamb_dissi}, amplitude damping has been recognized as an effective computational resource for sequential tasks of quantum reservoir computing. Specifically, these works show how a source of non-unital noise, as already theoretically suggested in Refs.~\cite {non_uni_prx, non-uni.barren} breaks the symmetry of the quantum evolution by biasing the state of the system away from the maximally mixed configuration. 
Previously, some of the authors have addressed learning algorithms by exploiting noise for both Boltzmann machines \cite{noe2024quantum} and quantum reservoir computing \cite{molteni,noiyqrcmonz}. Here, we propose an algorithm that induces damping, ensuring non-unital dynamics,  without relying on native hardware dissipation. In contrast with previous embodiments on intrinsically noisy quantum hardware \cite{noiyqrcmonz, kubota2023temporal}, such an algorithm, essentially based on controlled rotations of the accessible qubits, allows for a completely tunable damping rate, thus ensuring long-term computations. In particular, our method offers a tunable framework to explore the interplay between memory retention and non-unitality in the evolution of the reservoir. Our approach is validated through numerical simulations concerning the approximation of nonlinear functionals \cite{narma} and the prediction of chaotic time series~\cite {RC_mackey_glass}. Moreover, we investigate the effective utility of the quantum reservoir by showing the role of quantum correlations in the memory capacity of the network, in particular when dealing with encoding ansatzs characterized by different entangling gates \cite{gotting2023exploring}. Indeed, the dynamics of the system subject to mid-circuit measurements is described by quantum trajectories that encode an input-dependent evolution, allowing the reservoir to retain information about past inputs without the need for a classical memory buffer. Eventually, we test the algorithm on a currently available IBM superconducting quantum computer. It demonstrates improvements in the ability to memorize and process information over long time intervals compared to previous approaches based on intrinsic noise \cite{noiyqrcmonz, kubota2023temporal}, despite being implemented on newer, and thus less noisy, hardware. In conclusion, these results outline a robust strategy for scalable, noise-independent quantum reservoir computing that is compatible with fault-tolerant quantum devices. 

\begin{figure*}
    \centering
    \includegraphics[scale = 0.8]{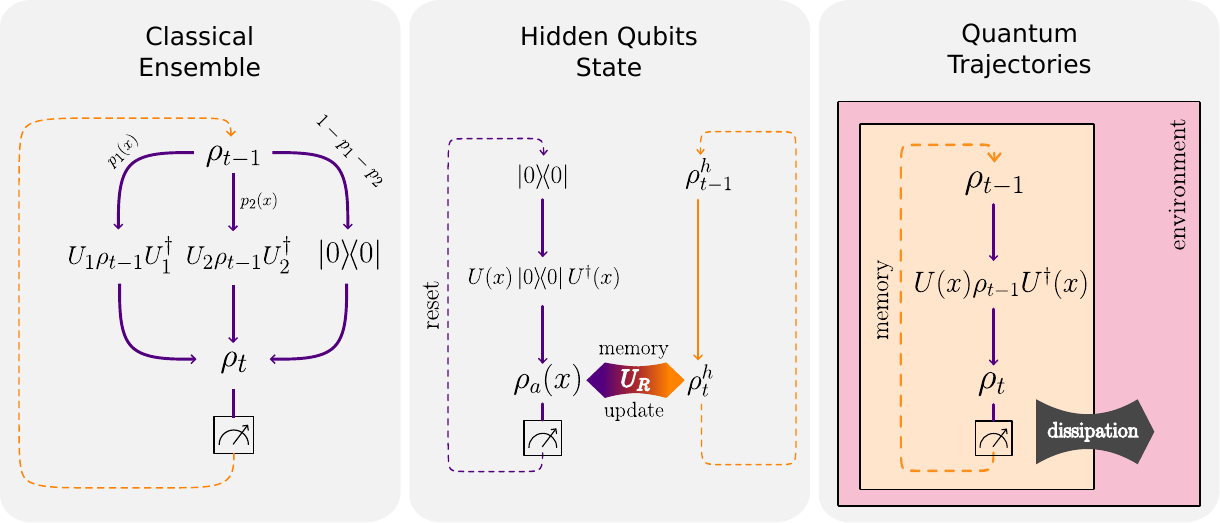}
    \caption{\textbf{Three paradigms for gate-based quantum reservoir computing}. In the first \cite{molteni,chen2} (left panel), the input data, denoted with $x$, is encoded by applying two fixed unitaries $U_1, U_2$ with classical probability depending on $x$. Moreover, a hard reset of the register is performed with fixed probability, ensuring the fading memory of the network. In the second \cite{hu2024overcoming,xiong2025fundamental} (central panel), the information is encoded in the state of a register of accessible qubits by some parametric unitary $U(x)$. After the input encoding, the accessible qubits interact with a hidden reservoir, which contains the past information processed by the network, through the unitary $U_R$, and then are measured to extract the processed data. Here, the non-unitality of the dynamics is ensured by the repeated reset of the accessible qubits. In the third \cite{kubota2023temporal, noiyqrcmonz} (right panel), the input value is again encoded in the state of the accessible qubits employing parametric unitaries. The whole register of employed qubits is measured and left to evolve in the collapsed state without  (orange dashed line). As shown in the work, the loss of information due to the repeated mid-circuit measurements must be prevented by some dissipation mechanism that ensures the non-unitality of the evolution.}
    \label{tav.1}
\end{figure*}
\section{Results}
In this Section, we present the main results of our work. Before that, we begin by summarizing the main concepts of gate-based quantum reservoir computing by distinguishing three possible approaches, which differ in terms of the strategy adopted to ensure memory retention. Next, we first demonstrate that induced dissipation compensates for quantum information loss, thus ensuring the persistence of the computational process. Then, we discuss the role of entanglement in the input encoding circuit by investigating the behavior of two- and three-qubit gates. Building on this a priori investigation, we verify the computational performance of the algorithm on two standard benchmarks. The first addresses the reconstruction of a nonlinear mapping, while the second aims at predicting a chaotic dynamical system.  Eventually, we test our algorithm on IBM superconducting quantum computers, showing that the controlled damping algorithm allows better performance compared to implementations that rely only on intrinsic noise \cite{noiyqrcmonz, kubota2023temporal}. 
\subsection{Gate-based quantum reservoir computing}
The general aim of sequential data processing with quantum reservoir computing consists in defining a mapping from an input space, constituted by temporal data $x  \in \mathcal{I} $, into another target time series $y \in \mathcal O$, through read-out operations of the information encoded in the reservoir. Generically, denoting the state of the reservoir with $\rho(t)$, defined on the Hilbert space $\mathcal{H}$ that describes a quantum system, the evolution of a quantum reservoir computer can be written as
\begin{align}
    \begin{cases}
        \rho_{t} = \mathcal T(\rho_{t-1}, x_{t})\\
        y_{t} = h(\rho_{t})
    \end{cases}\,.
\end{align}
where the time evolution of the system over discretized time intervals is expressed as $\dots\rho_{t-1},\rho_{t},\rho_{t+1}\dots$, $\mathcal T:\mathcal{S}(\mathcal{H}) \rightarrow \mathcal{S}(\mathcal{H})$ is any completely positive trace preserving (CPTP) channel that describes the dynamics of the quantum system, and $h$ is a suitable readout operation. Then the reservoir computer defines a causal functional $\mathcal{C}\colon \mathcal{I}\rightarrow \mathcal{O}$, which acts as
\begin{equation}
y_t = \mathcal C(x)_t = \mathcal{C}\left(x_{|_t}\right) = h\left(\mathcal{T}(\rho_{t-1},x_t)\right)
\end{equation}
where $x_{|_t} = (x_0,\dots,x_{t-1},x_t)$ indicate the input sequence truncated at time $t$. In particular, the training of a reservoir computer consists of finding the readout weights, encoded in the readout operation $h$, in order to minimize the discrepancy between the target functional $S(x)$ and the functional $\mathcal{C}$ induced by the dynamics. In principle, any quantum system may act as the reservoir, as long as the sequential input data can be effectively embodied in the evolution of the system and the reservoir dynamics adequately preserve the memory of past inputs. In particular, while the current state of the system must depend in a non-trivial way on the past values of the time series, it must also be able to give greater importance to more recent inputs progressively, thus achieving the so-called fading memory property \cite{boyd.chua}.
 Within the variety of possible approaches to quantum reservoir computing \cite{qrc_opp}, we develop the gate-based evolution of a quantum circuit of $N$ qubits as the computational reservoir. This choice is preferred among the three conceptually different approaches to gate-based quantum reservoir computing, also summarized in Fig. \ref{tav.1}. The first uses classical statistics over mixed states with engineered reset to balance memory \cite{chen2, molteni}. The second relies on an architecture with resettable accessible qubits and hidden memory qubits \cite{hu2024overcoming, xiong2025fundamental}, while the third — pursued in this work — operates without resets, leveraging quantum trajectories and controlled damping to maintain memory in fully accessible systems. We extend the discussion in the Appendix. In the algorithm exploited in this work, detailed in the Methods, the input encoding is performed through parametric unitary matrices. Concurrently, the qubits are measured online along the execution of the circuit and then allowed to evolve starting from the collapsed state. Repeated measurements of the quantum circuit pose an issue, since it is well known that they progressively induce information loss, eventually hindering the computational capabilities of the reservoir computer. Our work shows how the controlled damping of the qubits can overcome this issue, making the algorithm stable over time intervals significantly longer than the typical information survival time in the measured system. 
\begin{figure*}[t]
    \centering
     \includegraphics[scale = 0.8]{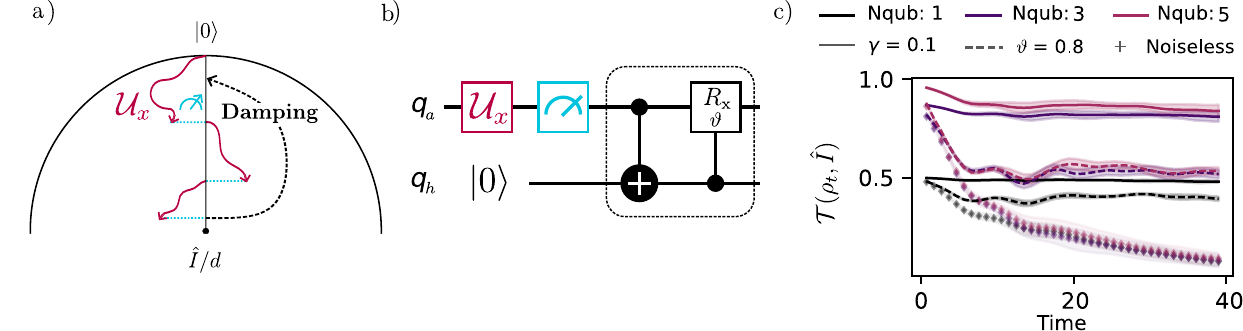}
    \caption{ \textbf{Impact of damping on decoherence of the reservoir. }{\bf a)} A sketch of the evolution of the state of a qubit on the Bloch sphere, taken from \cite{noiyqrcmonz}. The unitary $\mathcal U_x$ responsible for the input encoding moves the state of the reservoir qubit on a shell with fixed distance from the origin. After the measurement, the state collapses on the north-south axis, reducing the distance from the maximally mixed 
    state. {\bf b)} The circuit for the induced dissipation after input encoding and mid-circuit measurements for 1 qubit. The dissipation is induced by a CNOT gate followed by a controlled rotation of angle $\vartheta$. The same scheme applies to each qubit in the circuit. {\bf c)} Trace distance of the circuit from the $\hat{I}$ along 40 time steps of input encoding and mid-circuit measurements, averaged over 10 repetitions with random inputs. If the dynamics is affected by amplitude damping (dashed line) or by induced dissipation towards an ancilla register (continuous line), the state remains deviated from $\hat{I}$ during the evolution. }
    \label{tav.2}
\end{figure*}
\begin{figure*}[t]
    \centering
    \includegraphics[scale = 0.8]{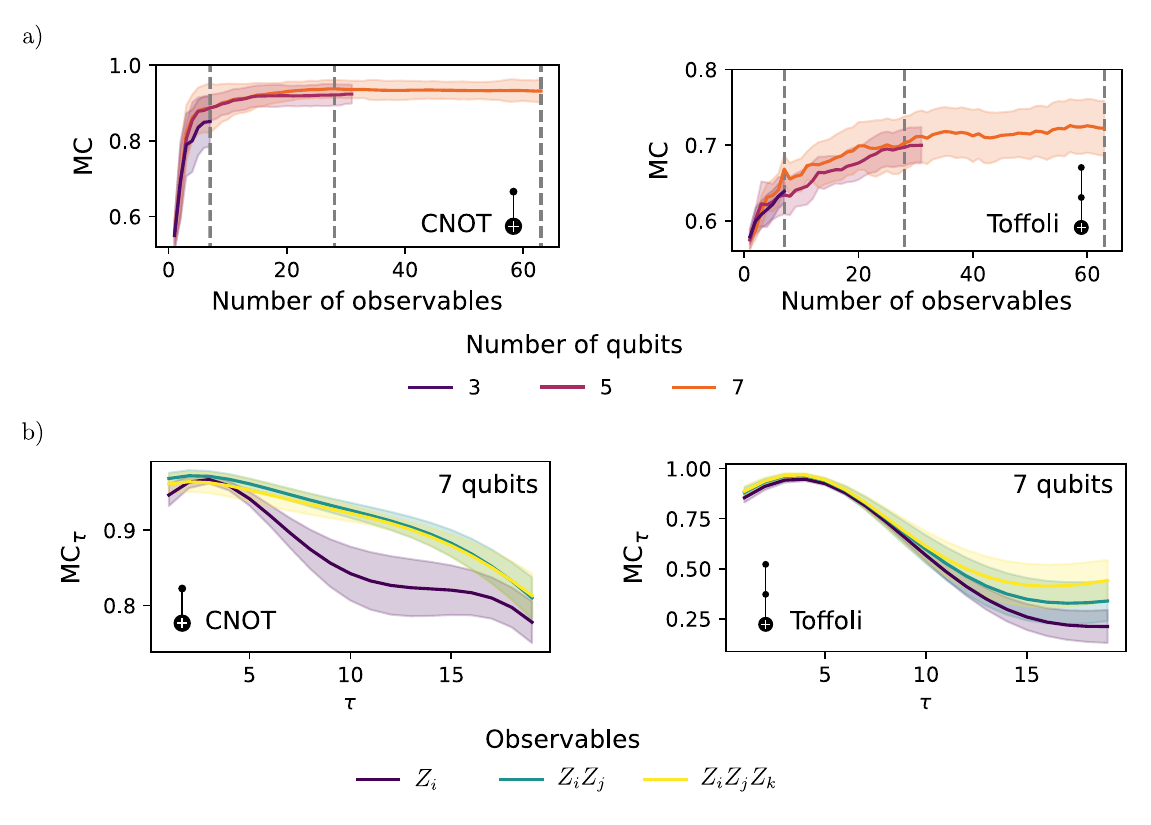}
    \caption{ \textbf{Memory capacity with different encoding unitaries.} {\bf a)} Memory capacity with $\tau_{\text{max}} = 20$ for increasing number of Z observables included in the readout, respectively for an input encoding with two and three-qubit entanglement. The dashed grey lines are for $N_{\text{obs}} = 7, 28, 63$ and correspond to the set of correlated observables involving 1,2, and 3 qubits, respectively, for a reservoir composed of 7 qubits. {\bf b)} Memory capacity for increasing time-span ($MC_\tau$) for the two different encoding anstatz. The three lines correspond to different readout observables. Purple line represents the caso of single ($N_{\text{obs}} = 7$) observables, while green and yellow lines correspond to adding double ($N_{\text{obs}} = 23$, and triple ($N_{\text{obs}} = 63$) correlated observables. The simulations employ an induced damping corresponding to $\vartheta = 0.8$.} 
    \label{tav.3}
\end{figure*}
\subsection{Information retention by induced dissipation}
We begin the presentation of our results by demonstrating how our algorithm prevents information loss during input processing, enabling sustained reservoir computing. In many gate-based quantum reservoir computing implementations \cite{noiyqrcmonz, kubota2023temporal, hu2024overcoming}, including the one presented in this work, the input data are sequentially injected into the qubits register, and, concurrently, decoded by repeated mid-circuit measurements on the computational basis. Therefore, given that no reset of the accessible qubits is performed, the state of the system is described by a mixture that encompasses all possible trajectories depending on the measurement outcomes, when considered as a statistical ensemble of many repetitions of the circuit. A key requirement for effective reservoir computing is that the system’s dynamics is sufficiently expressive to reliably separate distinct inputs \cite{martinez2025input}, such that the resulting states $\rho_x = \mathcal{U}_x \rho$ and $\rho_y = \mathcal{U}_y \rho$ are distinguishable after the input injection through some unitary matrix. Following Holevo–Helstrom theorem \cite{Helstrom1969,holevo1973bounds}, the distinguishability of two different quantum states is quantified by their trace distance, namely $\mathcal{T}(\rho,\sigma) = \frac{1}{2}\norm{\rho-\sigma}_1$ with $\norm{A}_1 = \text{Tr}\left(\sqrt{A\,A^\dag}\right)$. In particular, as discussed in Ref. \cite{noiyqrcmonz}, the trace distance of the internal states of the reservoir  corresponding to different values of the input satisfies the inequality
\begin{align}\label{continuity}
    \mathcal{T}\left(\rho_x - \rho_y\right) \le C(U)\cdot C(|x-y|)\cdot \mathcal{T}\left(\rho - \hat{I}/d \right) \,.
\end{align}
Here, $\rho$ is the initial state of the circuit before the injection of the input and $\hat{I}/d$ is the maximally mixed state,  with $d = 2^N$ denoting the dimension of the related Hilbert space.
The two constants, denoted with $C(U)$ and $ C(|x-y|)$,  depend on the unitary used for the input injection and the distance input values, respectively.  In particular, a larger deviation from $\hat{I}$ improves the separability of distinct inputs \cite{noiyqrcmonz}. As shown in Fig. \ref{tav.2}, the repeated mid-circuit measurements drive the internal state of the reservoir $\rho_t$ in the maximally mixed state, $\hat{I}/d$, in which the encoding of any input value is ineffective since $\mathcal{U}_x\,\hat{I} = \hat{I}$ for any value $x$. Indeed, any unitary rotation does not increase the distance from the identity, since
\begin{align}
\norm{\mathcal{U}_x \rho - \hat{I}/d}_{1} = \norm{\mathcal{U}_x( \rho - \hat{I}/d)}_{1} = \norm{\rho - \hat{I}/d}_1
\end{align}
while projective measurements decrease it, namely
\begin{align}
    \norm{\Pi \rho \Pi^\dag - \hat{I}/d}_1 =  \norm{\Pi (\rho - \hat{I}/d )\Pi^\dag }_1 \le \norm{\rho - \hat{I}/d }_1\,.
\end{align}
\paragraph*{\textbf{i. External dissipation.}}
The concentration of the reservoir in the maximally mixed state can be prevented by exploiting the action of a non-unital channel. We recall that a quantum channel $\mathcal E$ is non-unital when $\mathcal E (\hat{I}) \!= \hat{I}$. As an example, the amplitude damping channel can be adopted for this purpose, as it tends to relax the qubit state toward 
$\ket{0}$ with a decay rate $\gamma$ \cite{Domingo2023, kubota2023temporal, noiyqrcmonz}. We recall that amplitude damping can be easily modeled by the application of the quantum channel described by the Kraus operators $K_0 = \begin{pmatrix}
  1 & 0 \\
  0 & \sqrt{1-\gamma}
\end{pmatrix},\,\, K_1 = \begin{pmatrix}
  0 & \sqrt{\gamma}\\
  0 & 0
\end{pmatrix}$.  Although this approach offers significant advantages—being implementable on quantum computers with high dissipation rates—it may fail to deliver good performance, as it is hard to properly tune the noise using error mitigation techniques. While in principle it may be possible to enhance dissipation in many practical implementations of quantum computers, such as photonic or superconducting devices, it is usually difficult to design error mitigation techniques that specifically correct only the unital part of the noise, such as depolarizing and dephasing.
\begin{figure*}[t]
    \centering
    \includegraphics[scale = 0.8]{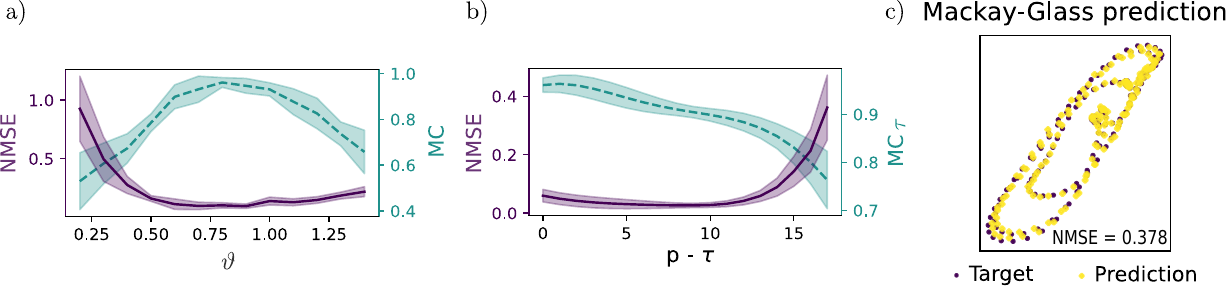}
    \caption{{\textbf{Memory capacity and benchmarks on a simulator. }\bf a)} Memory capacity and mean error in solving the NARMA task for increasing value of the damping, parametrized by $\vartheta$. Here, the NMSE is computed as the average of the NMSE for each NARMA-p task for $p\in(0,20)$. {\bf b)} Memory capacity and normalised error for NARMA task with increased time horizon. Namely, here NARMA is the mean error for each NARMA-p task with fixed $p$ running on the x-axis. The damping is fixed at the observed optimal value $\theta = 0.8$. {\bf c)} Prediction of the next step dynamics of the Mackay-Glass chaotic attractor. Only the test set is depicted. All the experiments in this Figure are conducted on a local simulator. In the Figure is reported the mean value of the error and an error band of 1 standard deviation.}\label{tav.4}
\end{figure*}
\begin{figure*}[t]
    \centering
    \includegraphics[scale = 0.85]{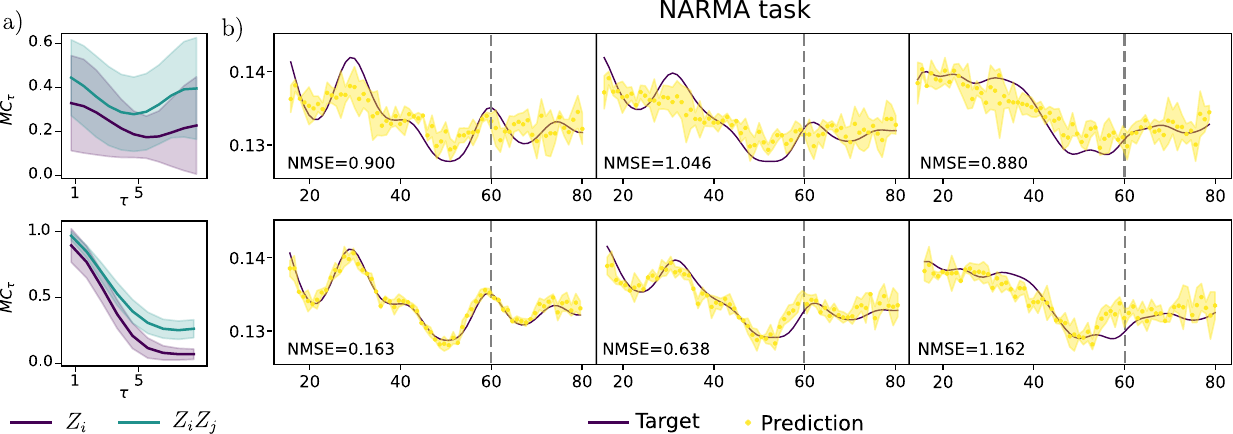}
    \caption{\textbf{Experiments on a superconducting quantum computer. }{\bf a)} Memory capacity for increasing time window without (above) and with (below) induced damping, corresponding to a fixed rotation angle $\vartheta = 0.6$. The solid line indicates the mean value over 5 executions, while the error band corresponds to 1 standard deviation. {\bf b)} Reconstruction of the NARMA-p task for, from left to right, $p = 2,5,8$. The line above contains results for an execution without induced damping, while the line below reports the results for a damping corresponding to $\vartheta = 0.8$. The experiment is run 5 times, and we represent the mean value of the prediction along with an interval of $1$ standard deviation. The grey dashed line divides the training and test sets in the learning algorithm. The washout period is not depicted.}\label{tav.5}
\end{figure*}
\paragraph*{\textbf{ii. Dissipation towards an ancilla register.}}
Alternatively, we propose a new approach by exploiting artificial dissipation toward an ancilla register to remove part of the superposition of the circuit. This methodology induces amplitude amplification of the zero state of the qubits in the reservoir, which balances the loss of quantum information and steers the state away from the maximally mixed state. Compared to external dissipation, the digital approach allows to perform the optimal tuning of the parameter that controls the dissipation. In this work, we induce the damping by connecting each qubit to an ancilla in the state $\ket{0}$ and a controlled rotation $CR_X(\vartheta)$ of a fixed angle $\vartheta$, then partially damping the state of the qubit as
\begin{align}
    \alpha \ket{0} + \beta \ket{1} \underset{CR_X(\vartheta)}{\rightsquigarrow} \left(\alpha + i \sin(\vartheta/2)\right) \ket{0} + \beta \cos(\vartheta/2) \ket{1}
\end{align}
after the input injection, as described in detail in Fig. \ref{tav.2}. We show in Fig. \ref{tav.2} that both methods, namely dissipation due to amplitude damping and that towards an ancilla, ensure that the internal state of the reservoir is kept away from the maximally mixed state, thereby guaranteeing the long-term operation of the algorithm, in contrast to what is observed in previous works \cite{rep, kubota2023temporal}. Indeed, the stability of the internal state of the reservoir in terms of its purity allows for an a-priori infinitely long computation.

\subsection{Input encoding and correlated observables}
After establishing an algorithm for the theoretical functioning of a gate-based echo state network, we now pass to investigate the effect of different encoding unitaries on its memory capacity. In particular, in this Section, we test two different input encoding ansatz, involving two-qubit and three-qubit logic gates, respectively. We show that, correspondingly, leveraging observables involving multiple qubits enhances the memory capacity of the reservoir. Indeed, adding entangling gates in the input encoding unitaries allows the system to exploit more computational degrees of freedom. As a proof of principle, we test two sets of unitaries, including interactions of two and three qubits, respectively, in addition to rotations that encode the current input value. In the first case, the input encoding $\mathcal{U}_x$ is composed of CNOT gates with bond dimension 1 and rotations $ R_X(x)$. On the other hand, three-qubit entangling gates are embodied by Toffoli gates CXX. In particular, the angles of the rotations are exploited for the encoding of the value of the input. The precise ansatz of the two circuits is detailed in the Methods section.  The reservoir signal is reconstructed from the expectation values of the Pauli-Z observables. In addition to the individual observables $Z_i,\, i = 1,\dots,N_a$,  we include the expectation values of joint observables involving multiple qubits, such as $Z_i\otimes Z_j,\, Z_i\otimes Z_j \otimes Z_k, \dots$, and so on. Notably, this does not require additional measurements of the system, since all of these observables commute. Yet, because joint observables are sensitive to correlations across multiple qubits, they enable the discrimination of entangled states that would otherwise be indistinguishable. This, in turn, truly allows for the exploitation of the system's quantum degrees of freedom, thus enforcing a potential quantum advantage in the reservoir computation. This point is clarified by a trivial example involving the separable two-qubit state $\ket{++}$ and an entangled one, $\ket{\phi^+}$. Indeed, while they are indistinguishable employing single observables, since $\langle Z_i\rangle_{\ket{++}} = \langle Z_i\rangle_{\ket{\phi^+}} = 0$ for $i = 1,2$, they are discriminated by joint observables, since $\langle Z_1 \otimes Z_2\rangle_{\ket{++}} = 0 $ but $\langle Z_1\otimes Z_2\rangle_{\ket{\phi^+}} = 1$. We tested the effect of joint observables on the memory capacity of the reservoir. In this test, as detailed in the Methods, the reservoir computer is trained via linear regression on the readout weights to reproduce a sequential input $x_t$ delayed in time by a fixed window $\tau$, namely $(\hat{y}^\tau)_t = x_{t-\tau}$. Then, the memory capacity (MC) is quantified by the amount of variance in the delayed input  $\hat{y}^\tau $ that can be recovered from the trained output $y^\tau$ \cite{jaeger2001short}, namely
\begin{equation}
\mathrm{MC}_\tau =  \frac{\mathrm{Cov}\left(y^\tau,\hat y^\tau\right)}{\mathrm{Var}(y^\tau)\mathrm{Var}(\hat y^\tau)}\,.
\end{equation}
Moreover, we quantify the overall memory capacity as \begin{align}
    \mathrm{MC} = \frac{1}{\tau_{\text{max}}} \sum_{\tau=0}^{\tau_{\text{max}}}\mathrm{MC}_\tau.
\end{align}
As expected, in the case of input encoding that uses only two-qubit gates, only the exploitation of two-qubit observables in the readout of the reservoir’s information significantly improves the memory capacity. On the other hand, observables that correlate three qubits provide benefits only in the case of input encoding with Toffoli gates. We summarize these results in Fig. \ref{tav.3}, where we show the memory capacity as a function of the number of $Z$ observables considered in the construction of the readout. The experiment is repeated for 12 different input instances of length $T = 200$ and three reservoir circuits consisting of 3, 5, and 7 qubits, respectively.

\subsection{Performance on benchmark tasks}
Building on previous insights, we test our algorithm on two benchmark tasks for reservoir computing. First, we investigate the memory capacity of the quantum echo state network by reconstructing the NARMA-p functional \cite{narma}. Moreover, we demonstrate the predictive capabilities of the network by forecasting the next step in the dynamics of a Mackey-Glass attractor in its chaotic regime. Both tasks are detailed in the Methods section. First, using a local simulator, we selected the optimal damping rate for the task under consideration. As a typical issue in reservoir computing, in which a strategy for a a-priori tuning of the hyperparameter often lacks, the best damping rate depends on both the task and the specific features of the circuit, and, currently, there is no general method to determine the optimal damping rate a priori. However, the network performs well on a sufficiently generic reconstruction task such as NARMA, precisely in the regime where it also exhibits the highest memory capacity. Thus, memory capacity serves as a reliable indicator of the network's performance. The results are reported in panels 1–2 of Fig.~\ref{tav.4}. Each experiment was repeated 12 times using a 7-qubit circuit. Implementation details can be found in the Methods section. We refer to the Supplementary Material for an in-depth analysis of the NARMA task, which allows for relating the absolute value of the NMSE reported in the paper to the actual accuracy in reconstructing the sequence. Moreover, as further evidence of the effectiveness of our algorithm, we employed the echo state network for the forecasting of chaotic dynamics. In this task, the reservoir computer receives at each step an input value taken from a time series governed by a differential equation and is required to output the next step in the sequence. The results of the experiment are reported in the panel $c$ in Fig.~\ref{tav.4}. 
\subsection{Experiment on a superconducting quantum computer}
Finally, we validate our algorithm by conducting experiments on a cloud-accessible superconducting quantum computer. Although these experiments are hindered by some limitations imposed by the remote computing service, such as the maximum number of gates executable in a single run, they show positive outcomes that improve the state of the art. Indeed, similar experiments were already conducted in previous papers \cite{kubota2023temporal, noiyqrcmonz}, achieving reasonable results for NARMA-2. Interestingly, these results were obtained employing older quantum hardware characterized by greater damping, which naturally enables the computation even in the absence of a specific algorithm for the amplitude amplification theoretically required along with repeated measurements. In this work, our algorithm effectively enables the execution of tasks requiring greater network memory, obtaining satisfactory results up to NARMA-5, while for larger memory intervals, the results progressively deteriorate. However, motivated by tests performed on an emulator that does not suffer from the limitations inherent to quantum hardware, we are confident that our algorithm allows the execution of tasks requiring the integration of information over longer periods. As shown in Fig. \ref{tav.5}, the induced damping drastically improves the memory capacity of the reservoir computer. Moreover, the observation on the beneficial effect of considering correlated observables along with two-qubit entangling gates is confirmed. In these experiments, we employ only CNOT gates in the input encoding unitary to better accomplish the limitation imposed by the remote service in terms of entangling gates, allowing for the execution of a reasonable number of shots of the circuit. The application of induced damping improves the network's performance, allowing the execution of the NARMA-p task with good results up to $p=5$. For harder tasks, the results begin to deteriorate due to quantum noise in the hardware. Indeed, while non-unitary noise such as amplitude damping enhances performance by contributing to the learning \cite{kubota2023temporal, noiyqrcmonz,Domingo2023}, other noise sources like dephasing and depolarizing progressively degrade the qubit state, limiting the memory capacity of the reservoir computer. 

\section{Methods}
\subsection*{Architecture of the gate-based echo state network}
We begin by describing the architecture of our gate-based echo state network, which employs an N-qubit circuit to encode and process time-dependent information. Then, a linear readout is trained by linear regession to replicate a mapping $S(u) = \hat y$ between an input sequence $u = \left\{u_t\right\}_{t=0,\dots, L}$ and a target sequence $\hat y = \left\{\hat{y}_t\right\}_{t=0,\dots, L}$. We denote with $\rho_t$ the density operator that describes the state of the N-qubit circuit at time $t$. Its one-step evolution $\widetilde{\rho}_{t+1} = \mathcal U_{x_{t+1}}(\rho_t)$ is given by a unitary evolution, which encodes the value of the current input $x_{t+1}$ in the angles of some rotations. Precisely, the two-qubit entangling circuit evolves under the application of the following gates, namely 
\begin{equation}
\begin{aligned}
 U(x) = \prod_{i=0}^{N_a - 1} \overline{U}_{i,i+1}\,,\\
\overline{U}_{i,i+1} = \text{CX}^{i,i+i}\left(\hat{I}^0 \otimes \hat{I}^1 \dots R_X^i(x)\otimes \dots \otimes Id^{N_a}\right)   
\end{aligned}
\end{equation}
where we are denoting with $\text{CX}^{i,i+i}$ the CNOT gate with $i^{\text{th}}$ and $i+1^{\text{th}}$ qubits of the circuit as control and target, respectively, and with $R_X^i(x)$ the rotation of the $i^{\text{th}}$ qubit with angle $x$. On the other hand, the ansatz characterized by 3-qubit gates is defined as
\begin{equation}
    \begin{aligned}
         U(x) = \prod_{i=0}^{N_a - 2} \overline{U}_{i,i+1,i+2}\,,\\
\overline{U}^3_{i,i+1} = \text{CCX}^{i,i+i,i+2}\left(\hat{I}^0 \otimes \hat{I}^1 \dots \otimes R_X^i(x) \otimes \dots \otimes Id^{N_a}\right)   
    \end{aligned}
\end{equation}
where $\text{CCX}^{i,i+1,i+2}$ is the Toffoli gate with qubit $i$ and $i+1$ as control, and $i+2$ as target. After the input encoding, the mid-circuit measurement of the register is performed on the computational basis and, as opposed to other approaches \cite{hu2024overcoming, xiong2025fundamental}, the system is let evolve in the collapsed state after the measurements. In this way, at each time step, the ensemble $\rho_t$, which describes the state of the circuit when considering the experiment repeated for many shots, is reminiscent of the sequential input $x_{t'}$ for any $t' \le t$. Formally, we have
 \begin{align}
     \rho_{t+1} = \Pi_N \widetilde{\rho}_{t+1} \Pi_N^\dag 
 \end{align}
 where, denoting with $m_{s,t+1}\in \left\{0,1\right\}$ the outcome of the measure of the $i^{\text{th}}$ qubits $i$ at time $t+1$, we are writing
 \begin{align}
     \Pi_N = \bigotimes_{i=1}^N|m_{i,t+1}\rangle\langle m_{i,t+1}| 
 \end{align}
  Concurrently, the reservoir signal at time $t$ is constructed online as the expectation value of the $Z_i$ Pauli operators and their correlated tensor products, namely \begin{equation}
     z_t = \left[\langle Z_1 \rangle_{\rho_t}, \dots, \langle Z_M\rangle_{\rho_t}, \dots, \langle Z_i\otimes Z_j \otimes Z_k \rangle_{\rho_t}\dots \right]^T \,.
 \end{equation}
 In the following Section, we first describe in detail the ansatz of the circuit and the training procedure that allows for reconstructing the wanted functional given the reservoir signal.
 \subsection*{The architecture of the two encoding circuits}
\begin{figure*}[t]
        \includegraphics[width = 1.0\textwidth]{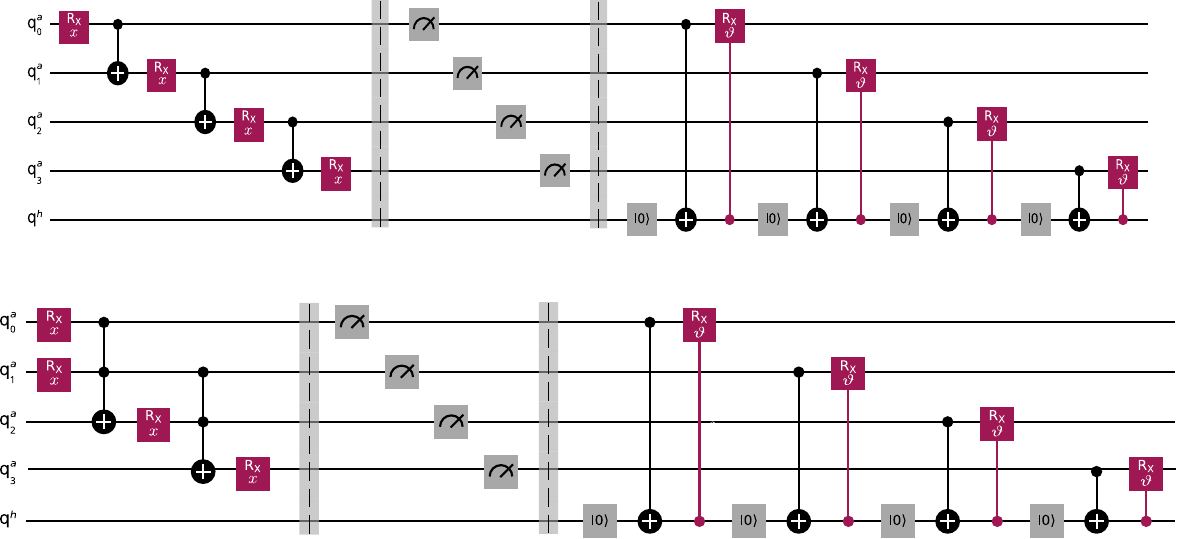}
    \caption{The circuits employed as the reservoir computer with two and three-qubit entangling gates, respectively. This block is repeated for each of the processed time series, exploiting mid-circuit measurements during execution. The ancilla register $q^h$ is employed for the induced dissipation. } \label{tavola06}
\end{figure*}

The two circuits employed in the experiment, which differ in the entangling gates in the input encoding, with CNOT and Toffoli gates, respectively, are reported in Fig. \ref{tavola06}. After the input encoding, the accessible qubits are measured online, and the system is left to evolve in the collapsed state. We remark that mid-circuit measurements are now available on many of the quantum hardware platforms currently available, and represent a key feature for the development of many quantum algorithms. After the measurements, the partial damping, driven by controlled rotation with a fixed parameter $\vartheta$, is induced by connecting the reservoir with an ancilla register. The same procedure is repeated for each value of the input sequential data. While the computation on a simulator benefits from the repeated reset of a single qubit, which essentially requires the multiplication of smaller matrices, in the case of implementation on quantum hardware, it is possible to leverage all available qubits to reduce the number of reset operations. This allows for a reduction in the computational time required for each reset and decreases operational error.

\subsection*{Training of linear readout}
We describe now the training procedure, which essentially reduces to a linear regression of the reservoir signal on the desired task. Indeed, reservoir computing aims to best reproduce a given nonlinear map $S(u) = \hat y$ by training a linear readout $W$. Namely, we aim to compute the optimal weights so that 
\begin{equation}
    W z_t = y_t  \simeq \hat y_t \,.
\end{equation}
After an initial washout period $T_{\text{wo}}$, required to erase the dependence on the initial state of the network and fulfill the so-called echo state property \cite{jaeger2001short}, the reservoir signal is stored in the matrix $H = \left\{z_t\right\}_{t=T_{\text{wo}} + 1,\dots, T_{\text{tr}}}$, where we are denoting with $\left( T_{\text{wo}}, T_{\text{tr}}\right)$ the time label of the training set of data. Then, the optimal readout weights, namely the ones for that $y_t = W z_t$ minimize the distance $\norm{\hat{y} - y}$ are computed by linear regression. Precisely, they are computed exploiting the Moore-Penrose pseudoinverse matrix, namely $ w = (H^T H)^{-1} H^T \hat y_{\text{tr}}$, where $\hat y_{\text{tr}} = \left\{\hat y _t\right\}_{t = T_{\text{wo}} + 1,\dots, T_{\text{tr}}}$ is the portion of target sequence used for the training. We refer to Ref.~\cite{molteni} for further details on the training algorithm. After the training, the predicted values of $y$ are given by 
\begin{equation}
    y_t = z_t \cdot W \qquad \text{for any } t \in [T_{tr} + 1, L] \,.
\end{equation}
Interestingly, the reservoir can be trained to fulfill different tasks that share the same input, since the internal dynamics is independent from the training algorithm on a specific task, thus reducing the computational cost of the execution. Moreover, as explained in the results, this allows for comparison of the intrinsic memory of the network -- with the memory capacity task -- and the performance on benchmark nonlinear tasks, as the NARMA task. 
\subsection{Benchmark tasks}
Here, we formally describe the two tasks employed in the experiments to validate the algorithm. They represent two typical benchmarks in reservoir computing implementations. In particular, the first evaluates the network’s ability to reconstruct past inputs, while the second assesses its predictive capabilities of a chaotic dynamical system.
\paragraph*{\textbf{i. NARMA task}}
The Nonlinear Auto-Regressive Moving Average dynamics (NARMA) task requires reproducing a nonlinear functional with past dependence in the sequential input, testing how well a reservoir computer can handle progressively longer dependencies. Formally, we define the NARMA-$p$ functional as
\begin{equation}
    \hat y_{t} = \alpha\, y_{t-1} + \beta\, \left(\sum_{l=0}^{p-1}y_{t-l}\right) + \gamma\, u_{t-p+1} u_t + \delta \,.
\end{equation}  
with $\alpha, \beta, \gamma, \delta = (0.3, 0.05, 1.5, 0.1)$. The input is sampled from the periodic function 
\begin{equation}
    u_t = 0.1 \sin{\left(2 \pi\frac{  2.11 t}{T}\right)}\sin{\left(2 \pi\frac{ 3.73 t}{T}\right)}\sin{\left(2 \pi\frac{  4.11 t}{T}\right)}\,.
\end{equation}
In the work, we fix $T = 200$ and we employ an input sequence with length 200 and 80 for the experiment on the local simulator and the remote quantum hardware, respectively. The test set has lengths of 130 and 50, while the first 20 steps are used for the washout of the initial condition of the reservoir. This choice is required to fulfill the computational limitations of the remote quantum service made available by IBM. We run simulations with $50.000$ shots, which represents a valid regime for approximating the expected values deterministically. The experiments on quantum hardware involve 10800 repetitions of the circuit, divided into 120 batches with 60 shots each.
\paragraph*{\textbf{ii. Mackey-Glass task}}
We test the predictive capabilities of the network by forecasting the next-step evolution of the Mackey-Glass equation in the chaotic regime. In the Mackey-Glass model, the dynamics is driven by a time-delay differential equation that exhibits chaotic behavior for certain values of the parameter $\tau$. Formally, the equation reads 
\begin{equation}
\frac{dx}{dt} = \beta \frac{x(t - \tau)}{1 + x(t - \tau)^n} - \gamma x(t),
\end{equation}
where we fix $\beta = 0.2$ and $\gamma = 0.1$ the positive constants that control the growth and decay rates, respectively, and $\tau$ is the delay parameter. For small values of $\tau$, the system exhibits stable fixed points or periodic oscillations. For increasing values of $\tau$, the system exhibits a transition into chaotic dynamics. Here, we fix $\tau = 17$ in the experiments, well within the range of the chaotic regime. Then, the task is defined as the next-step prediction, namely 
\begin{equation}
    \hat y_t = x(t+1) \,,
\end{equation}
 In the experiment, we consider an input sequence of length 2000, sampled from the continuous time series with a fixed period of length $10^{-2}$. In both cases, the accuracy of the prediction is quantified by the normalized mean square error (NMSE), expressed as
\begin{equation}
\mathrm{NMSE}(y,\hat{y})= \frac{\sum_{t=T_{\text{tr} + 1}}^{t=L}(\hat y_t - y_t)^2}{\sum_{t=T_{\text{tr} + 1}}^{t=L} (\hat y_t - \mu) ^2}
\end{equation}
where $y$ is the vector of the predicted values after training, $\hat y$ is the target sequence, and $\mu = L - T_{\text{tr}} \sum_{t=T_{\text{tr} + 1}}^{t=L} \hat y_t$ is the mean value of the target sequence. The training and the test set have lengths of 1300 and 500 steps, respectively. 
\subsection*{Experimental settings}
For the simulations, we used a 36-core Intel Xeon processor (4.3 GHz) with 128 GB of RAM.  In experiments involving quantum hardware, we used the IBM\_TORINO quantum processing unit, which mounts a Heron1 processor, equipped with 133 qubits. We access it through the IBM cloud service with an on-demand subscription.
\section{Conclusion}
We implemented a gate-based echo state network utilizing a qubit circuit for processing sequential data, proposing an algorithm that ensures the effectiveness of the dynamics of the quantum reservoir. Indeed, damping has already been recognized as a necessary resource for enabling the functioning of a gate-based echo state network \cite{kubota2023temporal}, as it ensures the separability of inputs over long-term computations \cite{noiyqrcmonz}. In particular, the loss of information due to repeated mid-circuit measurements is prevented by inducing the damping of the accessible qubits. In this work, we proposed a ad-hoc ansatz involving controlled rotations that, in contrast with previous approaches based on intrinsic noise in the quantum hardware, ensures the full controllability of the damping rate. Our algorithm enables an unbounded processing of information, limited only by the capabilities of the supporting hardware, without requiring a reset of the accessible register \cite{hu2024overcoming}. Moreover, unlike previous approaches that rely on the intrinsic noise of the hardware to ensure non-unital dynamics \cite{kubota2023temporal, noiyqrcmonz}, the explicit control of the induced damping allows for optimizing the dissipation-induced fading memory according to the nature of the task at hand. Furthermore, building up on our algorithm, we highlight the computational utility of the degrees of freedom associated with the state space of the quantum reservoir, which are fully exploited by encoding the data into entangled states and subsequently using correlated observables for the readout of the information processed by the reservoir, which, in turn, improves the network’s memory capabilities. Eventually, we test the method on standard benchmarks of reservoir computing, numerically and with experiments on a cloud-accessible superconducting quantum hardware. These experiments demonstrate the feasibility of our algorithm on currently available quantum hardware by improving the state of the art in terms of the memory capacity and the expressiveness of the echo state network in reconstructing nonlinear functionals that require a longer integration time. Besides that, our algorithm is suitable for future implementation on faultless quantum computers, as it is not based on intrinsic dissipation, unlike previously proposed algorithms \cite{kubota2023temporal,noiyqrcmonz}.

\section*{Acknowledgements} 
E.R. and E.P. are supported by Qxtreme project funded by the Partenariato Esteso FAIR
(grant N. J33C22002830006). F.M. and E.P. are supported by PRIN-PNRR PhysiComp (grant N. G53D23006710001). L.N. is partially funded by ENI S.p.A.
\section*{Code and data availability}
The data and the code are made available by the authors upon reasonable request.
\section*{Competing interests} Authors declares no competing interests.
\section*{Authors contribution}
E.R. and F.M. designed the algorithm, the computational framework and conceived the experiments. E.R E.R, F.M. and L.N. analysed the data and curated the visualization. 
E.R. and F.M. wrote the first of the manuscript with
input from all authors. All the authors contributed to the final version of the manuscript. E.P. conceived the study and were in charge of overall direction and planning.
 \medskip
\bibliography{bib}
\bibliographystyle{naturemag}

\clearpage
\appendix
\renewcommand{\thefigure}{A-\arabic{figure}}
\setcounter{figure}{0}
\onecolumngrid  

\section{Three families of gate-based quantum reservoir computers}
We discuss here three different families of gate-based echo state networks for sequential data processing. These approaches differ in their strategies for input encoding and memory management, while all rely on exploiting the expectation values of certain observables as readout functions. 
\subsection*{i. Gate-based QRC with classical statistics.}
The first proposal of a gate-based quantum reservoir computer involved the input encoding in the classical statistics of a mixture state \cite{chen2}. Precisely, the density operator that describes the system is left evolved according to the equation,
\begin{align}
    \rho_{t} = (1- \epsilon) \big( x_{t} \mathcal{E}_0 + (1-x_{t})\mathcal{E}_1\big)\rho_{t-1} + \epsilon\, \sigma 
\end{align}
where $\mathcal{E}_0, \mathcal{E}_1$ are two fixed CPTC channels, and $\epsilon \in (0,1)$ is the reset rate to the fixed state $\sigma$. Usually, the quantum channels $\mathcal{E}_0, \mathcal{E}_1$ are embodied by a circuit of fixed unitary transformations, while $\epsilon$ is optimized to balance the memory of the system with its fading memory \cite{molteni}. 
\subsection*{ii. Gate-based QRC with reset of accessible qubits.}
A second possible approach that exploits the degrees of freedom of the quantum system as the computational reservoir involves the distinction between $N_a$ accessible qubits, used for input encoding, and $N_h$ hidden qubits, acting as the memory storage \cite{hu2024overcoming, xiong2025fundamental}. Correspondingly, the state of the system can be written as $\rho = \rho^a \otimes \rho^h$, where we are highlighting the accessible part of the register, used for information encoding and decoding, and the hidden nodes, responsible for the memory of the recurrent network. In this setting, at each time step, the current value of the input data is injected into the accessible register by exploiting parametric unitary evolution of some fixed initial state,
\begin{align}
    \rho_t^a = U(x_t)\rho^a_*\, U^\dag(x_t) \equiv \mathcal{U}_{x_t} \rho_*^a\,.
\end{align}
Here and below, we are denoting with $\mathcal{U}_x \rho =  U(x)\rho\, U^\dag(x) $ the unitary evolution of $\rho$ under the parametric unitary $U(x)$. Typically, in practical implementations, $\rho^a_*$ is chosen as the state after qubits resetting, namely $\rho^a_* = \ket{0}\bra{0} ^ {\otimes N_a}$. This is compatible with the allowed reset operation on quantum hardware. After the input injection, the accessible qubits are left to interact with the hidden layer of the reservoir, which acts as the memory container, according to some fixed dynamics $U_R$ that entangles accessible and memory qubits, namely 
\begin{align}
    \rho_{t} = U_R \left(\rho_t^a \otimes \rho_{t-1}^h \right)U_R^\dag \,,
\end{align} 
where $\rho_{t-1}^h$ denotes the density operator that describes the hidden nodes of the reservoir. 
In this way, the information about the past data of the input, contained in the state of the reservoir of hidden qubits, is retrieved through the dynamics induced by $U_R$. Eventually, the information is collected in the expected value of some observables $\langle O \rangle_{\rho_t^a}$ of the accessible qubits, and they are reset to allow for the next state preparation.
\subsection*{iii. Gate-based QRC without reset of accessible qubits.}
A third possible approach, which is also pursued in this work, involves dealing with a fully accessible reservoir of $N_a$ qubits \cite{rep, noiyqrcmonz}, without any reset of the qubits, and thus exploiting quantum trajectories as a memory retention mechanism. While, as in the previous case, the input encoding is performed through parametric unitary matrices, in this family of algorithms, the qubits are measured online along the execution of the circuit and then allowed to evolve starting from the collapsed state. Namely, one can write the evolution of the system as
\begin{align}\label{tre}
\rho_t = \Pi \big( U(x_t)\rho_{t-1}\, U^\dag(x_t) \big) \Pi^\dag \equiv \Pi \big( \mathcal{U}_{x_t}\rho_{t-1}\big)  \Pi^\dag
\end{align}
where $\Pi$ represents the action of a projective measurement on the computational basis. Thus, information retention relies on the statistics of the observables of the accessible qubits only, without the need for a hidden reservoir, as described in the following Sections. Nonetheless, repeated measurements of the quantum circuit pose an issue, since it is well known that they progressively induce information loss, eventually hindering the computational capabilities of the reservoir computer. Our work shows how the controlled damping of the qubits can overcome this issue, making the algorithm stable over time intervals significantly longer than the typical information survival time in the measured system. 
\section{An explicit computation for a 1 qubit circuit} \label{trace_norm.app}
We describe in this Appendix the evolution of a 1-qubit reservoir circuit, highlighting the depolarization induced by repeated measurements and how this is prevented by the induced damping. We recall that the algorithm procedure consists of applying a unitary rotation followed by a projective measurement on the computational basis. Since we do not consider the measurement outcome in subsequent evolution, and we repeat the computation for many shots, the overall result is described by a statistical mixture of the computational basis states \( \ket{0} \) and \( \ket{1} \). Formally, we can write 
\begin{equation}
    \rho_{t} \rightarrow 
   \Pi 
    U \rho_t U^\dagger
 \Pi^\dagger \equiv \rho_{t+1} \,.
\end{equation}
For simplicity, we assume the unitary is a rotation about the x-axis.  Thus, fixing the current state density matrix as $
\rho_t = 
\begin{pmatrix}
\rho_{00} & 0 \\
0 & 1 - \rho_{00}
\end{pmatrix},
$ and applying the encoding rotation $R_x(u) = 
\begin{pmatrix}
\cos\left(\frac{u}{2}\right) & -i \sin\left(\frac{u}{2}\right) \\
-i \sin\left(\frac{u}{2}\right) & \cos\left(\frac{u}{2}\right)
\end{pmatrix}$, the evolved state of the reservoir after the projective measurement is given by 
\begin{equation}
    \rho_{t+1}
    = 
    \begin{pmatrix}
    \rho_{00} \cos^2\left(\frac{u}{2}\right)
    + (1 - \rho_{00}) \sin^2\left(\frac{u}{2}\right) & 0 \\
    0 &
    (1 - \rho_{00}) \cos^2\left(\frac{u}{2}\right)
    + \rho_{00} \sin^2\left(\frac{u}{2}\right)
    \end{pmatrix}. 
\end{equation}
In other words, the measurement operation eliminates all off-diagonal terms, i.e., the coherence terms. As a result, the overall evolution can be described by a decoherence map, 
\begin{equation}
    \rho_{t+1} = (1 - \gamma) \rho_{t} + \gamma(\hat{I}- \rho_{t}) 
    = \rho_{t}(1 - 2\gamma) + \gamma \hat{I} 
    = (1 - p) \rho_{t} + \frac{p}{2} \hat{I}\,,
\end{equation}
where we are denoting with 
$\gamma = \sin\left(\frac{u}{2}\right)$, and with $p = 2\gamma$ the rate of decoherence. \begin{figure*}[h]
    \centering
    \includegraphics[scale = 1.0]{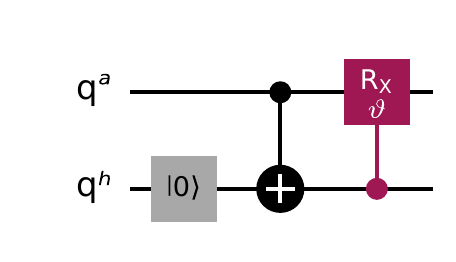}
    \caption{The gates for the induced damping of the accessible qubit. It is connected to an ancilla qubit initialised in $\ket{0}$ with a CNOT, and then rotated with a controlled rotation of angle $\vartheta$, which allows for the tuning of decay.}
    \label{app.fig.1}
\end{figure*}
In particular, applying rotations in the upper hemisphere of the Bloch sphere, thus restricting $u \in \left[0, \frac{\pi}{2}\right] $, we have \( \gamma \in \left[0, \frac{1}{2}\right] \), and therefore \( p \in [0, 1] \). Thus, the repeated mid-circuit measurements act as a depolarizing channel, progressively driving the system toward the maximally mixed state \( \hat{I}/2 \). Similar calculations lead to the same result for an N-qubit register. \\ 

We now describe the ansatz of the circuit that prevents depolarization by inducing artificial damping in the qubit. Writing the state of accessible qubit $q^a$ after the encoding rotation and the projective measurement as $\rho^a = |\alpha|^2 \ketbra{0}{0} + |\beta|^2 \ketbra{1}{1}$, where we are denoting with $\alpha = \alpha(u)\,$ and $ \beta = \beta(u)$ some coefficients that depends on the unitary rotation, we entangle it with an ancilla qubit $q^a$ initialised at $\ket{0}$ with a CNOT gate, namely
\begin{equation}
      \rho = q^a \otimes q^h=  |\alpha|^2 \ketbra{0,0}{0,0} + |\beta|^2 \ketbra{1,1}{1,1}.
\end{equation}
Then, we induced the damping of the accessible qubit by a controlled rotation $\text{CR}_X(\vartheta)$, with control on $q^a$ and target on $q^a$. The damped state is written as
\begin{equation}
\rho = |\alpha|^2 \ketbra{0,0}{0,0} + |\beta|^2 ( c_\vartheta \ket{1,1} + d_\vartheta\ket{0,1})(c_\vartheta^* \bra{1,1} + d_\vartheta^* \bra{0,1})
\end{equation}
where are we denoting with
$
c_\vartheta = \cos\left(\frac{\vartheta}{2}\right), $ and $
d_\vartheta = -i \sin\left(\frac{\vartheta}{2}\right)
$
the coefficients that quantify the damping. Indeed, tracing out the ancilla qubit, the reduced density matrix of the accessible qubit is 
\begin{equation}
\begin{aligned}
\rho_{q_0} &= \operatorname{Tr}_{q_1}(\rho) 
= |\alpha|^2 \ketbra{0}{0} + |\beta|^2 d_\vartheta^2 \ketbra{0}{0} + |\beta|^2 c_\vartheta^2 \ketbra{1}{1}\,.
\end{aligned}
\end{equation}
As a result, this procedure transfers part of the population of state \(\ket{1}\) into state \(\ket{0}\), damping the state of the qubit from $|\alpha|^2$ to $|\alpha|^2 + c_\vartheta^2 |\beta|^2$.

\section{The problem of overfitting}
\begin{figure*}[h]
    \centering
    \includegraphics[scale = 0.52]{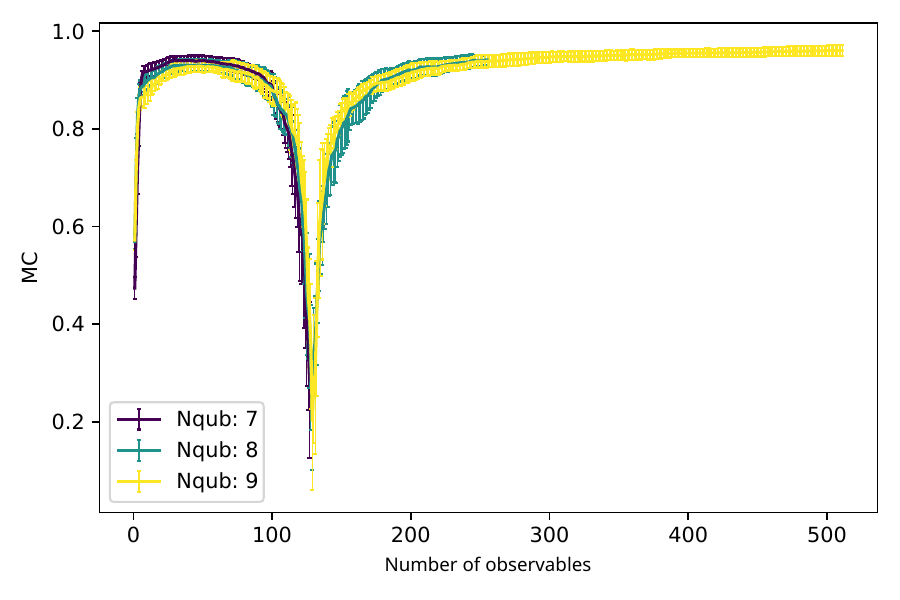}
    \caption{Memory capacity of the network for an increasing number of observables. We note the overfitting in the training, which emerges when the number of observables matches the length of the training set ($N\sim 130$).}
    \label{app.fig.3}
\end{figure*}

We discuss here the problem of overfitting in training the linear readout, which arises when the number of parameters to be trained is approximately equal to the number of features in the training data, and we show that it is solved when the number of observables is increased. Denoting with $N_{\text{obs}}$ the number of observables used as readout and with $L$ the length of the training set of data, we recall that the training essentially consists of solving the linear system
\begin{equation}
    W  \cdot H = y  \simeq \hat y \,.
\end{equation}
where $\hat y$ represents the target data, $y$ the computed output after the training, $W$ is the vector of the $N_{\text{obs}}$ trainable parameters, and $H$ is the $N_{\text{obs}} \cross L$ matrix that contains the reservoir signal-- i.e., the value of the expected values of the measured observables -- throughout the evolution of the circuit.  When the system is overdetermined ($L \gg N$), it avoids the overfitting of the model. As expected, we note that when $L\sim N_{\text{obs}}$, the training completely fails, leading to almost zero memory capacity of the network, when the number of observable matches the length of the training set, which is fixed at 130, as shown in Fig. \ref{app.fig.3}. However, the effectiveness of the network is recovered by increasing the number of observables,  thus allowing the implementation of the algorithm on a larger set of qubits, which indeed would require a larger set of observables as readout. We remark that, as exploited in our work,  increasing the number of commuting observables does not require further measurement of the system. In the experiment discussed in the Main Text, we restrict the number of observables in the overdetermined regime $L \gg N$.
\section{More details on NARMA task }
We report here the detailed representation of the NARMA-p task until $p = 18$ in order to relate the value of NMSE with the precision of the reconstruction. The results are obtained employing a local simulator with $50000$ shots. Such a number of shots, beyond which significant statistical variations are no longer detectable, allows for a reliable approximation of the deterministic expected value of the observables. For $p\le 9$ the algorithm reaches almost perfect reconstruction of the nonlinear functional. For increasing values of $p$, corresponding to a larger window of integration for the execution of the task, some noise around the mean value of the target sequence is visible. Nonetheless, the overall trend of the dynamics is preserved, and the reservoir computer is able to effectively predict the peaks and valleys of the target sinusoidal function. This shows that the reservoir retains its functionality over time, remaining unaffected by the loss of coherence induced by mid-circuit measurements.
\begin{figure*}[h]
    \centering
    \includegraphics[width = 1.0\textwidth]{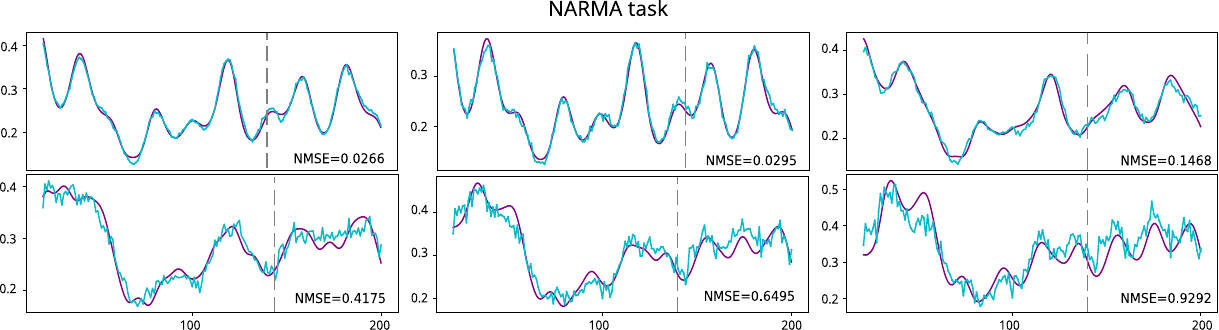}
    \caption{The NARMA-p task for $p = 3,6,9,12,15,18$ from top left to bottom right. The purple line and light blue line represent the target and task sequence, respectively. The grey dashed line separates the test and training sets. }
    \label{app.fig.4}
    \end{figure*}

\end{document}